\documentclass[11pt]{article}
\usepackage{DGfest,epsfig}
\usepackage{times}

\def\be{\begin{equation}}
\def\ee{\end{equation}}
\def\bea{\begin{eqnarray}}
\def\eea{\end{eqnarray}}

\def\e{\mathrm{e}}

\def\re{\mathop{\mathrm{Re}}}
\def\im{\mathop{\mathrm{Im}}}

\def\d{\mathrm{d}}

\def\Re{\mathrm{Re\:}}
\def\Im{\mathrm{Im\:}}

\begin{document}
\baselineskip 11.5pt
\title{EXTENDED DERIVATIVE DISPERSION RELATIONS}

\author{R.F. \'AVILA}

\address{Instituto de Matem\'atica, Estat\'{\i}stica e
Computa\c c\~ao Cient\'{\i}fica\\
Universidade Estadual de Campinas, 13083-970 Campinas, SP, Brazil}

\author{ M.J. MENON }

\address{Instituto de F\'{\i}sica Gleb Wataghin\\
Universidade Estadual de Campinas, 13083-970 Campinas, SP, Brazil}

\maketitle\abstracts{It is shown that, for a wide class of functions with physical 
interest
as forward scattering amplitudes, integral dispersion relations
can be replaced by derivative forms without any high-energy approximation.
The applicability of these extended derivative
relations, in the investigation of forward
proton-proton and antiproton-proton elastic scattering,
is exemplified by means of a Pomeron-Reggeon model with totally
nondegenerate trajectories.}

\centerline{\textit{Contribution to "Sense of Beauty in Physics"}}
\centerline{\textit{Miniconference in Honor of Adriano Di Giacomo on his 70th Birthday}}
\centerline{\textit{Pisa, Italy, Jan. 26-27, 2006}}

\section{Introduction}

Dispersion relations constitute a traditional and important 
mathematical tool in several
areas of physics, not only as a formal theoretical result, 
but also as a powerful phenomenological 
framework. In particular, for \textit{elastic} hadron-hadron
scattering, analyticity, unitarity and crossing lead to dispersion
relations, which connect the real and the imaginary parts of the  amplitude
as function of the energy, allowing  simultaneous investigation
of particle-particle and
antiparticle-particle scattering.
Originally introduced in integral forms~\cite{idr}, these
\textit{Integral Dispersion Relations} (IDR)
have two kinds of limitations: their nonlocal character (in order to
evaluate the real part, the imaginary part must be known over
all the integration space) and the restricted class of functions that allows
analytical integration. The first limitation (and, in part, the second too)
can be avoided by means of \textit{Derivative Dispersion Relations} (DDR), which have been established
only for the region of high and asymptotic energies~\cite{ddr}. In a recent 
work~\cite{am04} 
we present a critical review on the replacement of
IDR by DDR and also references to outstanding results.
However, despite the  important developments provided by this
derivative approach, the high-energy
condition (specifically, center-of-mass energies above 10 - 20 GeV) turns out
difficult any attempt to perform global fits to the experimental data,
connecting information from low and high energy regions.

In a previous paper~\cite{am05} we have shown that, for a class of functions
of physical interest as forward elastic scattering amplitudes, the IDR 
can be replaced by derivative forms without the high-energy approximation,
which we call \textit{Extended Derivative Dispersion Relations} (EDDR). 
In this communication we first  review the replacement
of IDR by the EDDR and then discuss a novel example on the practical
applicability of both forms in the context of a Pomeron-Reggeon model,
with totally nondegenerate secondary trajectories. 
In Sec. 2 we recall the main formulas
related with the IDR and the standard DDR (high-energy approximation assumed).
In Sec. 3 we shortly review the replacement of IDR by the EDDR and then discuss
the applicability of these dispersion relations in 
simultaneous description
of the experimental data on the total cross section  and the ratio $\rho$ 
of the real to imaginary parts of the forward amplitude, from
proton-proton ($pp$) and antiproton-proton ($\bar{p}p$) scattering.
The conclusions are the contents of Sec. 4.

\section{Integral and Derivative Dispersion Relations}

For an elastic process, $m + m \rightarrow m + m$, in the forward direction,
IDR are expressed in terms of a crossing symmetric
variable, which corresponds to the energy of the incident particle
in the laboratory system, $E$.
For elastic $pp$ and $\bar{p}p$ scattering, polynomial boundedness
demands one subtraction and 
for crossing even ($+$) and odd ($-$) amplitudes, in the
physical region ($E:m \rightarrow \infty$), the IDR read~\cite{idr}

\begin{equation}
\re F_{+}(E)=  \frac{2E^{2}}{\pi}P\!\!\!\int_{m}^{+\infty}
\!\!\!\d E'
\frac{1}{E'(E'^{2}-E^{2})}\im F_{+}(E'),
\label{eq:1}
\end{equation}

\begin{eqnarray}
\re F_{-}(E)=  \frac{2E}{\pi}P\!\!\!\int_{m}^{+\infty} \!\!\!
\d E'
\frac{1}{(E'^{2}-E^{2})}\im F_{-}(E'),
\label{eq:2}
\end{eqnarray}
where we have omitted the subtraction constant.

Basically, at high energies, the replacement of IDR by DDR is analytically 
performed~\cite{ddr,am04}
by considering  the  limit $m \rightarrow 0$ in Eqs. (1) and (2) .       
Expansion of the integrand and then integration term by term lead to the 
standard DDR~\cite{ddr}

\begin{equation}
\Re F_{+}(E)= 
E\tan\left[\frac{\pi}{2}\frac{\mathrm{d}}{\mathrm{d}\ln E} \right]
\frac{\Im F_{+}(E)}{E},
\label{eq:bksp}
\end{equation}

\begin{equation}
\Re F_{-}(E)
=\tan\left[\frac{\pi}{2}
\frac{\mathrm{d}}{\mathrm{d}\ln E} \right]
\Im F_{-}(E).
\label{eq:bksi}
\end{equation}
Conditions on the convergence of the
above tangent series will be specified in what follows.

\section{Extended Derivative Relations}

\subsection{Analytical Results}

Let us consider the even amplitude, Eq. (\ref{eq:1}). Details on the
calculation can be found in our previous work~\cite{am05}; here we only summarize
the four main steps: (a) integration of Eq. (1) by parts;
(b) change of variable~\cite{cms}, $E \rightarrow m \e^{\xi}$\ ;
(c) expansion of the integrand in power series; (d)
integration term by term, under the assumption of uniform convergence of
the series associated with the function

\begin{eqnarray}
\frac{\mathrm{d}}{\mathrm{d}\ln E} \frac{\Im F(E)}{E}.
\end{eqnarray}

With this procedure and returning to the variable $E$ we obtain

\begin{eqnarray}
\Re F_+(E)&=&
- \frac{E}{\pi}\ln\left|\frac{m-E}{m+E}\right|\frac{\Im F_+(m)}{m}
\nonumber \\
&+&\frac{4E}{\pi}\sum_{p=0}^{\infty}\sum_{k=0}^{\infty}
\frac{\d^{2k+1}}{\d(\ln E)^{2k+1}}\left(\frac{\Im F_{+}(E)}{E}\right)
\frac{1}{(2p+1)^{2k+2}}
\nonumber \\
&+&\frac{2E}{\pi}
\sum_{k=0}^{\infty}\sum_{p=0}^{\infty}
\frac{\d^{k+1}}{\d (\ln E)^{k+1}}
\frac{\Im F_+(E)}{E}\frac{(-1)^{k+1}\Gamma(k+1,(2p+1)\xi)}{(2p+1)^{k+2}k!} \nonumber
\end{eqnarray}
which can be put in the final form

\begin{eqnarray}
\Re F_{+}(E)=
E\tan\left(\frac{\pi}{2}\frac{{\mathrm{d}}}{{\mathrm{d}}\ln 
E}\right)\frac{\Im F_+(E)}{E}
+\Delta^+(E,m),
\label{eq:newp}
\end{eqnarray}
where the factor $\Delta^{+}$ is given by

\begin{eqnarray}
& &\Delta^{+}(E,m)=
-\frac{E}{\pi}\ln\left|\frac{m-E}{m+E}\right|\frac{\Im F_+(m)}{m}\nonumber\\
&+&\frac{2E}{\pi}
\sum_{k=0}^{\infty}\sum_{p=0}^{\infty}
\frac{\d^{k+1}}{\d (\ln E)^{k+1}}
\frac{\Im F_+(E)}{E}\frac{(-1)^{k+1}\Gamma(k+1,(2p+1)\ln(E/m))}{(2p+1)^{k+2}k!}.
\nonumber
\end{eqnarray}

With analogous procedure for the odd relation we obtain

\begin{eqnarray}
\Re F_{-}(E)=
\tan\left(\frac{\pi}{2}
\frac{{\mathrm{d}}}{{\mathrm{d}}\ln E}\right)\Im F_-(E)
+\Delta^-(E,m),
\label{eq:newi}
\end{eqnarray}
where

\begin{eqnarray}
& &\Delta^-(E,m)=
-\frac{1}{\pi}\ln\left|\frac{m-E}{m+E}\right|\Im F_-(m)\nonumber\\
&+&\frac{2}{\pi}
\sum_{k=0}^{\infty}\sum_{p=0}^{\infty}
\frac{\d^{k+1}}{\d (\ln E)^{k+1}}
\Im F_-(E)\frac{(-1)^{k+1}\Gamma(k+1,(2p+1)\ln(E/m))}{(2p+1)^{k+2}k!}.
\nonumber
\end{eqnarray}

Equations (6) and (7) are the novel 
EDDR, which are valid, in principle, for any energy
above the physical threshold $E > m$. We note that the
factors $\Delta ^\pm \rightarrow 0$ as $E \rightarrow \infty$,
leading, in this case, to the standard DDR, Eqs. (3)
and (4).

Necessary and sufficient conditions for the convergence of the
tangent series have been established by Kol\'a\v{r} and Fischer~\cite{kf},
in particular through the following theorem:

\newtheorem{guess}{Theorem}
\begin{guess}
Let $f: R^1 \rightarrow R^1$. The series

\begin{eqnarray}
\tan\left[\frac{\pi}{2} \frac{d}{dx} \right] f(x)
\nonumber
\end{eqnarray}
converges at a point $x \in R^1$ if and only if the series
\begin{eqnarray}
\sum_{n=o}^{\infty} f^{(2n + 1)}(x)
\nonumber
\end{eqnarray}
is convergent.
\end{guess}

Since this Theorem insures the uniform
convergence of the series expansion associated with (5),
the above condition defines the class of functions for
which the EDDR hold. Other conditions are discussed by Kol\'a\v{r} and 
Fischer~\cite{kf}.

\subsection{Applicability in Elastic Hadron Scattering}

An important practical use of the derivative relations concerns simultaneous 
investigations on the 
total cross section (Optical Theorem) and the ratio $\rho$ 
of the real to imaginary parts of the forward amplitude. In terms of the 
symmetrical variable
$E$ these physical quantities are given, respectively, by 

 \begin{equation}
\sigma_{\mathrm{tot}}
=\frac{4\pi}{\sqrt{E^2-m^2}} \Im F(E,\theta_{\mathrm{lab}}=0),
\qquad
\rho(E) = \frac{\re F(E,\theta_{\mathrm{lab}}=0)}{\im F(E,\theta_{\mathrm{lab}}=0)},
\label{eq:5}
\end{equation}
where $\theta_{\mathrm{lab}}$ is the scattering angle in the laboratory system.

In order to check the consistences between the IDR and the EDDR in
an specific practical example, we consider, as a framework, a Pomeron-Regge
parametrization for the scattering amplitude, in which all the associated secondary reggeon
contributions are nondegenerate. In this case,
for $pp$ and $\bar{p}p$ scattering, the even ($+$) contributions comes from
the  $a_2$ and $f_2$ trajectories and the odd ($-$) contributions
from the $\rho$ and $\omega$ trajectories; the full parametrization also
includes a simple pole Pomeron contribution:

\begin{equation}
\Im F(E)=XE^{\alpha_{\tt I\!P}(0)}+Y_{a_2}E^{\alpha_{a_2}(0)}
+Y_{f_2}E^{\alpha_{f_2}(0)}
+\tau 
\left[Y_{\rho}E^{\alpha_{\rho}(0)}+Y_{\omega}E^{\alpha_{\omega}(0)}\right] ,
\end{equation}
where $\tau=+1$ for $pp$ and $\tau=-1$ for $\bar{p}p$. As usual, the Pomeron
and the Reggeon intercepts are expressed by

\begin{equation}
\alpha_{\tt I\!P}(0)=1+\epsilon, \qquad
\alpha_{i}(0)=1-\eta_{i},
\end{equation}
where $ i = a_2$, $f_2$, $\rho$ and $\omega$.

The point is to treat simultaneous fits to the
total cross section and the $\rho$ parameter from $pp$
and $\bar{p}p$ scattering and compare the results obtained with IDR, EDDR
and also standard DDR.
Schematically, with parametrization (9-10) for $pp$ and $\bar{p}p$
we determine $\Im F_{+/-}(E)$
through the usual definitions

\begin{equation}
F_{+} = \frac{F_{pp} + F_{\bar{p}p}}{2},
\qquad
F_{-} = \frac{F_{pp} - F_{\bar{p}p}}{2}
\end{equation}
and then $\Re F_{+/-}(E)$ by means of the IDR,
Eqs. (1-2), DDR, Eqs. (3-4) and EDDR, Eqs. (6-7). Returning to Eq. (11) we obtain
$\Re F_{pp}(E)$ and $\Re F_{\bar{p}p}(E)$ and,  at last,  Eq. (8) 
leads to the analytical connections between $\sigma_{\mathrm{tot}}(E)$ and
$\rho(E)$ for both reactions.

For $\sigma_{\mathrm{tot}}$ and $\rho$, we have compiled all the experimental data available
above the physical threshold~\cite{pdg}. The fits were performed through the CERN-Minuit code,
using the variable $s = 2(m^2 + mE)$.
However, with the present model (intended for the high-energy region), the large 
number of experimental points just above
this threshold allows reasonable statistical results only for an energy cutoff
at  $\sqrt s_{\mathrm{min}} =$ 3 GeV. 
The numerical results and statistical information on the fits are displayed in Table 1 
and the corresponding curves together with the 
experimental data are shown in Fig. 1.

\begin{table}[t]
\caption{Simultaneous fits to $\sigma_{\mathrm{tot}}$ and $\rho$,
from $pp$ and $\bar{p}p$ scattering, for
$\sqrt s_{\mathrm{min}} =$ 3 GeV
(308 data points), using Integral Dispersion Relations (IDR),
standard Derivative Dispersion Relations (DDR) and the Extended
Derivative Dispersion Relations (EDDR)\label{tab1}}
\vspace{0.4cm}
\begin{center}
\begin{tabular}{|c|ccc|}
\hline
& IDR & DDR & EDDR  \\
\hline
$X$ (mb)         & 1.6634  $\pm$ 0.0093  & 1.7586  $\pm$ 0.0080  & 1.6634  $\pm$ 0.0093\\
$Y_{a_2}$ (mb)   & -16.779 $\pm$ 0.038   & -19.600 $\pm$ 0.055   & -16.779 $\pm$ 0.038\\
$Y_{f_2}$ (mb)   & 20.792  $\pm$ 0.038   & 23.063  $\pm$ 0.054   & 20.792  $\pm$ 0.038\\
$Y_{\rho}$ (mb)  & -0.334  $\pm$ 0.038   & 0.441   $\pm$ 0.056   & -0.334  $\pm$ 0.038\\
$Y_{\omega}$ (mb)& -2.087  $\pm$ 0.049   & -2.587  $\pm$ 0.056   & -2.087  $\pm$ 0.049\\
$\epsilon$       & 0.08869 $\pm$ 0.00065 & 0.08402 $\pm$ 0.00061 & 0.08869 $\pm$ 0.00065\\
$\eta_{a_2}$     & 0.37679 $\pm$ 0.00091 & 0.5873  $\pm$ 0.0014  & 0.37679 $\pm$ 0.00091\\
$\eta_{f_2}$     & 0.37681 $\pm$ 0.00074 & 0.5389  $\pm$ 0.0010  & 0.37681 $\pm$ 0.00074\\
$\eta_{\rho}$    & 0.334   $\pm$ 0.014   & 0.574   $\pm$ 0.047   & 0.334   $\pm$ 0.014\\
$\eta_{\omega}$  & 0.700   $\pm$ 0.014   & 0.5746  $\pm$ 0.0085  & 0.700   $\pm$ 0.014\\
\hline
$\chi^2$         & 478.8                 & 405.7                 & 478.8\\
$\chi^2/F$       & 1.61                  & 1.36                  & 1.61 \\
\hline
\end{tabular}
\end{center}
\end{table}

These results demonstrate the complete equivalence between the IDR and the EDDR; moreover,
the high-energy approximation (DDR) indicates a slower increase for the total cross section
at the highest energies than that obtained with the exact results (IDR and EDDR)
and different behavior for $\rho(s)$ at low energies. We note
that the high values for $\chi^2/F$, in all the cases, are consequences of the particular
model considered (intended for high energies) and the fact that we have neglected the
subtraction constant as a free fit parameter. The important role played by this
parameter is discussed in our previous work~\cite{am05}.

\section{Conclusions}

We have obtained novel analytical expressions for the derivative
dispersion relations, without 
high-energy approximations.
These EDDR are intended for
any energy above the physical threshold and their applicability is restricted to the
class of functions specified by Theorem 1.
However, since the experimental data on the total cross sections
indicate a smooth variation with the energy (and a smooth 
systematic increase above
$\sqrt s \approx 20$ GeV), this class includes the majority of
functions of physical interest.
Using as a framework a Pomeron-Reggeon model without degenerate trajectories, 
we have demonstrated the numerical equivalence between the results obtained
with the IDR (finite lower limit $m$) and the EDDR.

\section*{Acknowledgments}
It is our pleasure to dedicate this work to Prof. Adriano Di Giacomo, on the
occasion of his 70th birthday. M.J.M. is deeply grateful to
Prof. Di Giacomo for all the support and encouragement, since 1987,
in particular for the hospitality at the Universit\`a di Pisa (1991 - 1993).
\textit{Parab\'ens, caro Di Giacomo e muitas felicidades!}

We are thankful to the organizers for the invitation
to contribute to this Volume and to
FAPESP for financial support (contracts No. 03/00228-0 and No. 04/10619-9).

\begin{figure}
\epsfig{figure=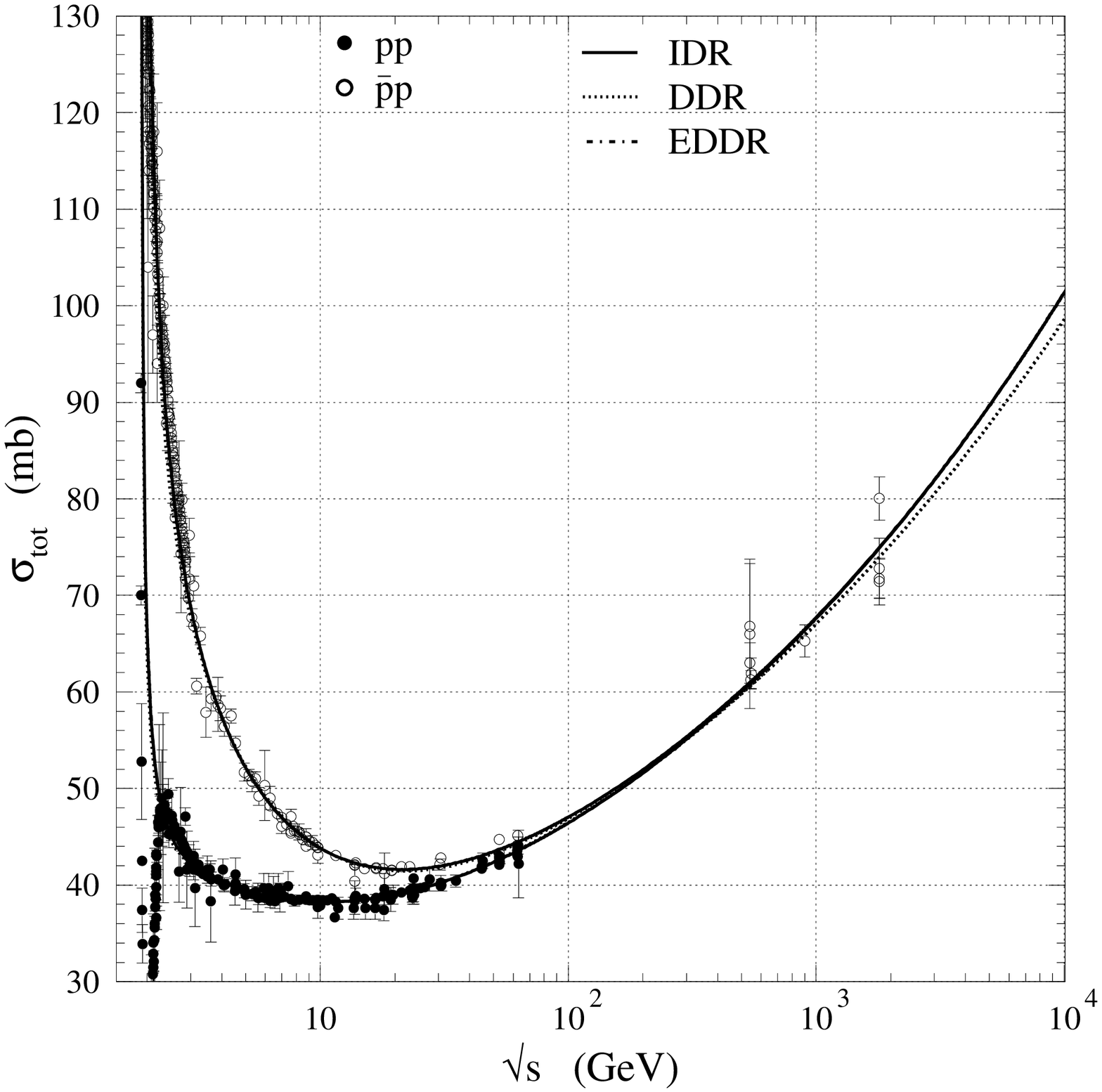,width=2.6in,height=3.58in}
\epsfig{figure=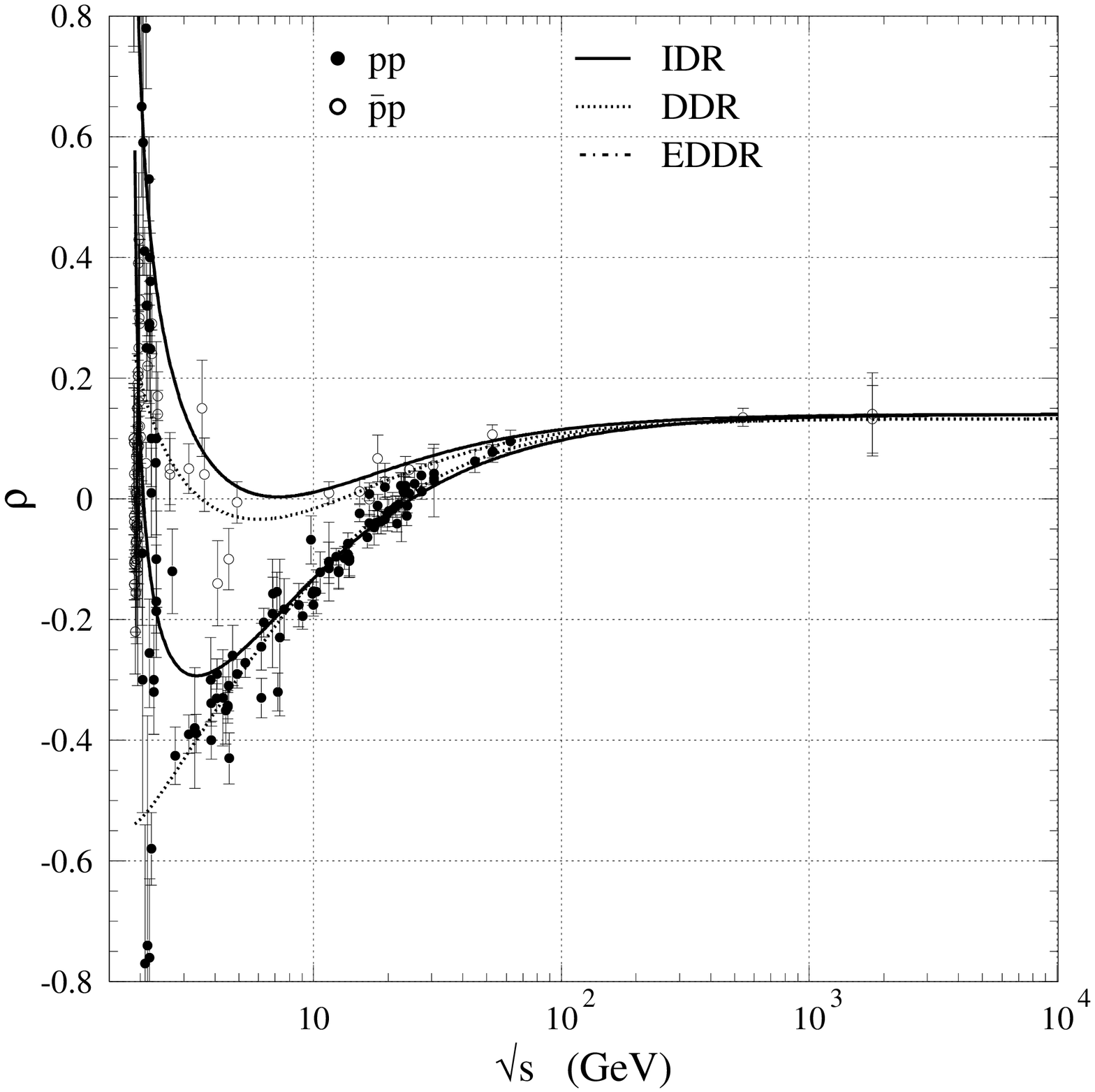,width=2.6in,height=3.58in}
\caption{Results of the simultaneous fits to 
$\sigma_{\mathrm{tot}}$ and $\rho$ from $pp$ and $\bar{p}p$ scattering.
\label{tcs}}
\end{figure}

\end{document}